\documentclass[sigconf,edbt]{acmart-edbt2023}

\def\BibTeX{{\rm B\kern-.05em{\sc i\kern-.025em b}\kern-.08em
    T\kern-.1667em\lower.7ex\hbox{E}\kern-.125emX}}

\usepackage{booktabs} % For formal tables

\setcopyright{rightsretained}

\acmDOI{}

\acmISBN{978-3-89318-088-2}

\acmConference[EDBT 2023]{26th International Conference on Extending Database Technology (EDBT)}{28th March-31st March, 2023}{Ioannina, Greece}
\acmYear{2023}

\author{Patrick Marcel, Ver\'onika Peralta, Faten El Outa}
\affiliation{%
  \institution{University of Tours}
  \streetaddress{}
  \city{} 
  \state{} 
  \postcode{}
}
\email{firstname.lastname@univ-tours.fr}

\author{Panos Vassiliadis}
\orcid{}
\affiliation{%
  \institution{University of Ioannina}
  \streetaddress{}
  \city{} 
  \state{} 
  \postcode{}
}
\email{panos.vassiliadis@cs.uoi.gr}

\title{A declarative approach to data narration}

\usepackage{graphicx}

\newcommand{\patrick}[1]{\textcolor{red}{#1}}
\newcommand{\vero}[1]{\textcolor{black}{#1}}

\newcommand{\faten}[1]{\textcolor{teal}{#1}}

\newcommand{\eop}{\hfill $\Box$}                       % end of example

\begin{document}

\begin{abstract}
This vision paper lays the preliminary foundations for
Data Narrative Management Systems (DNMS), systems
that enable the storage, sharing, and manipulation of
data narratives.
We motivate the need for such formal foundations
and introduce a simple logical  framework
inspired by the relational model.
The core of this framework is a Data Narrative
Manipulation Language inspired by the extended relational algebra.
We illustrate its use via examples and
discuss the main challenges for the
implementation of this vision.
\end{abstract}

\maketitle

\section{Introduction}

%Introduce data narratives: 
A data narrative (DN)
is a structured composition of messages that (a) convey findings over the data, and, (b) are typically delivered via visual means in order to facilitate their reception by an intended audience \cite{DBLP:conf/er/OutaFMPV20}.
Data narration \cite{DBLP:journals/tvcg/SegelH10,DBLP:journals/dagstuhl-reports/CarpendaleDRH16} refers to the notoriously tedious process of
crafting a DN by
extracting insights from data and telling stories with the goal of "exposing the unanticipated" \cite{DBLP:books/lib/Tukey77} 
and facilitating the understanding of insights. Data narration is practiced in many domains and by various domain experts, ranging from data journalists to public authorities.

\begin{figure}[ht]
    \centering
    \includegraphics[width=8cm]{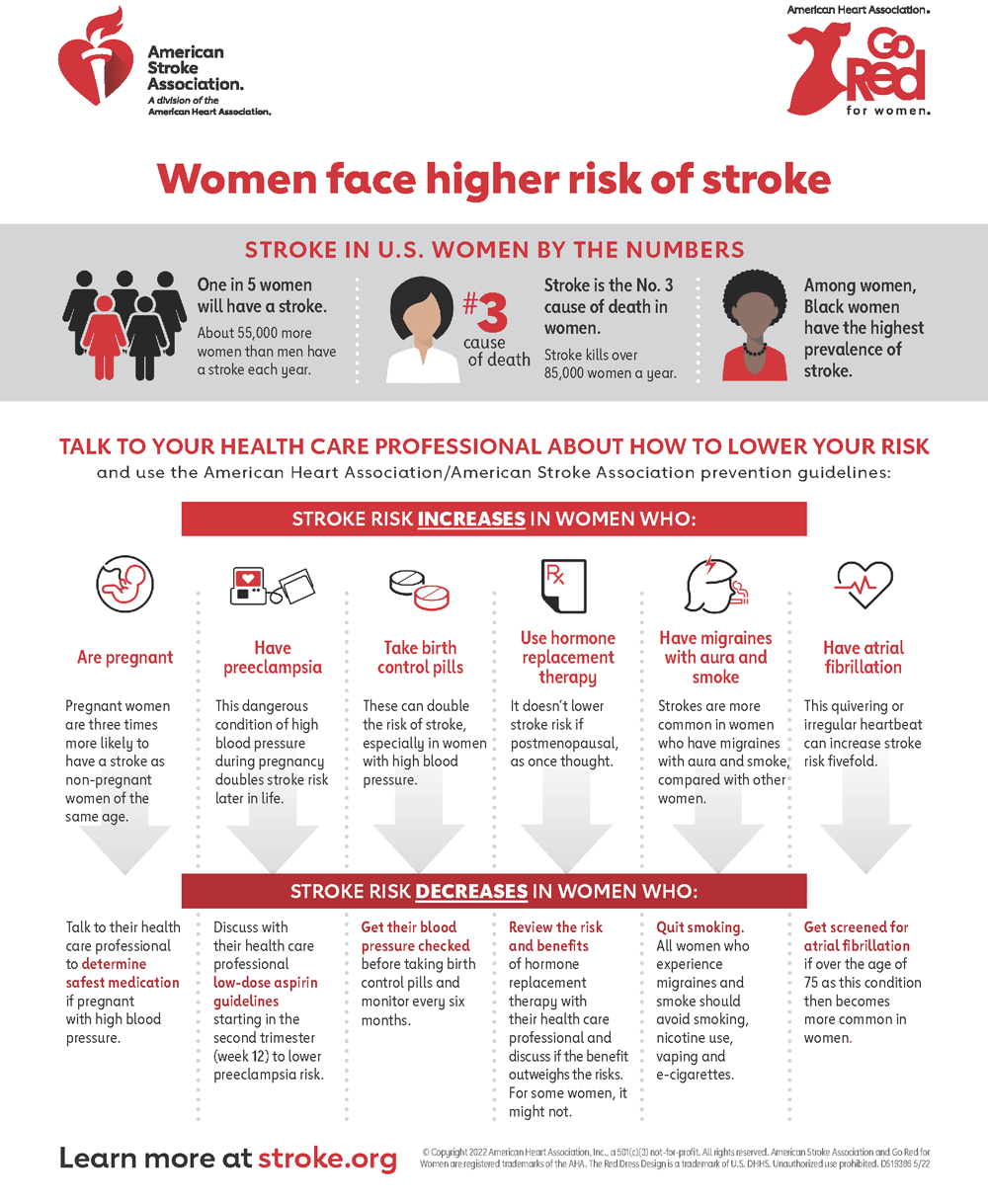}
    \includegraphics[width=8cm]{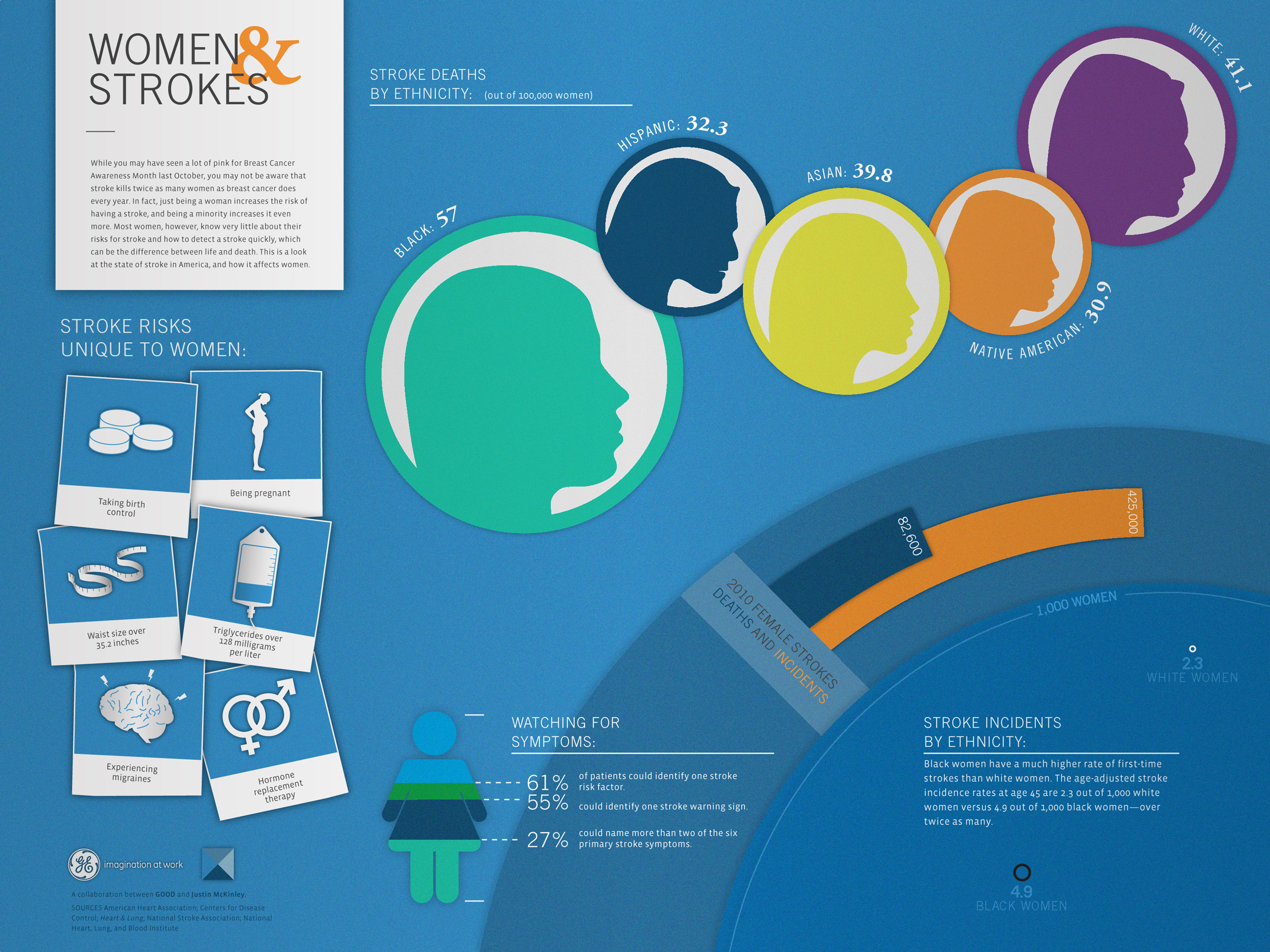}
    \caption{Examples of data narratives}
    \label{fig:2DN}
\vspace{-0.2cm}
\end{figure}

Consider the two infograhpics of Figure \ref{fig:2DN}\footnote{American Heart Association: "Women and Risk of Stroke Infographic", https://www.goredforwomen.org/en/know-your-risk/risk-factors/risk-of-stroke-in-women-infographic}$^,$\footnote{Good: "Infographic: Stroke, a Silent Killer of Women, Facts About Women and Strokes", https://www.good.is/infographics/facts-about-women-and-strokes}.
These two infographics can be seen as  "physical" representations
of the \vero{DNs.} %data narrative. 
In \cite{DBLP:conf/er/OutaFMPV20}, a conceptual model for \vero{DNs} %data narrative 
was proposed.
The goal of the present paper is to propose a logical representation
of \vero{DNs,} %data narratives, 
to bridge the gap between the physical and conceptual representations.

Let us first briefly review the proposed conceptual model for DNs
\cite{DBLP:conf/er/OutaFMPV20}.
% below from adbis
This model is based on 4 layers
following Chatman's organisation  \cite{chatman1980story}, who defined narrative as a pair of (a) \textit{story} (content of the narrative), and, (b) \textit{discourse} (expression of it). 
In the conceptual  model, the \textit{factual} layer handles the \textit{exploration} of facts (i.e., the underlying data) for
%via a set of \textit{collectors} that allow for manipulating facts with varied tools and 
fetching \textit{findings}
while the \textit{intentional} layer models the subjective substance of the story, 
identifying the \textit{messages}, \textit{characters} and \textit{measures} the narrator 
intends to communicate.
%and tracing how they are obtained through \textit{analytical questions}, according to an \textit{analysis goal}.
As to the discourse, the \textit{structural} layer models the structure of the \vero{DN,} %data narrative,  
its \textit{plot} being organized in terms of %\textit{acts} and  
\textit{episodes},
while the \textit{presentational} layer deals with its rendering,
that is communicated to the audience through visual artifacts 
%(\textit{dashboards}\footnote{We use the term  dashboard since it is general enough to accommodate various types of visualizations (e.g. a Business Intelligence dashboard, an infographics, a section in a python notebook, a section in a blog or web page).} and 
named \textit{dashboard components}).
The interested reader is redirected to \cite{DBLP:conf/er/OutaFMPV20}
for a deeper presentation of the model.

For instance, in the DN of Figure \Ref{fig:2DN} \vero{bottom,} %left,
an episode of the DN is rendered in the upper right
dashboard component, with the message indicating
that measure \vero{`stroke deaths'} %STROKE DEATHS BY ETHNICITY 
is 57/100000
for character `black women'.

\begin{figure*}[ht]
    \centering
    \includegraphics[width=14cm]{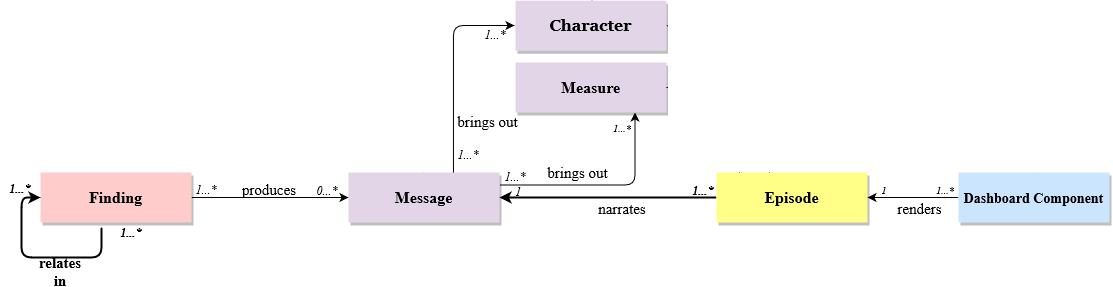}
    \vspace{-0.5cm}
    \caption{Selection of concepts used for the logical layer}
    \label{fig:subCM}
%\end{figure}
\end{figure*}

Figure \ref{fig:subCM} is an excerpt of this conceptual model \vero{considered} %, that is under consideration    %%% TO WIN 1 LINE
in the present paper.
Indeed, in this vision paper, our goal is to show the 
benefit of manipulating data narratives declaratively,
i.e., with a formal logical data model and 
a manipulation language. We voluntarily keep this
model and language simple, mostly inspired by the 
relational data model and extended relational algebra \cite{DBLP:books/daglib/0020812}.
As will be discussed in Section \ref{sec:discussion},
taking into account all the concepts
of the domain will need to revisit the model and language,
while keeping  the flavor of the manipulation
described here. 
These manipulations rely on the concept of \textit{message} 
which is the conceptual model's corner stone.
A  message is rooted in the facts analyzed,
conveying essential findings that can be related to one another.
The message allows introducing episodes, the building blocks of the discourse. 
Each episode of the discourse is specifically tied to a message which it aims to convey,
with dashboard components being their presentational counterparts.
%The relationship between messages and episodes is the basis for structuring stories that address analysis goals, narrated by structured discourses (with cohesive acts being the backbone of the narrative structure) and dashboards %(organized sub-areas of the visual presentation) 

%\begin{figure}[ht]

While the Web abounds with DNs, 
manipulating them 
in a declarative way
using the concepts
of this model
has not yet been proposed, to our knowledge. 
This  paper aims at filling this gap,
by envisioning a DN Management System (DNMS),
the foundations of which should include
a logical layer enabling the declarative 
 manipulation of DNs.

The outline of the paper is the following.
Section \ref{sec:motivation} motivates the 
 need for a logical layer.
Section \ref{sec:model} introduce the logical model for DNs
and Section \ref{sec:language} the algebra for manipulating DNs.
Section \ref{sec:examples} illustrates the languages
\vero{while} %with 
Section \ref{sec:rw} presents related work and
Section \ref{sec:discussion} concludes this vision paper
%with 
by discussing  the main challenges for the
implementation of \vero{DNMS.} %DN management systems.

\section{Motivation}\label{sec:motivation}

% \begin{figure*}[ht]
%     \centering
%     \includegraphics[width=8cm]{figures/womenstroke1.jpeg}
%     \hspace*{1cm}
%     \includegraphics[width=8cm]{figures/Stroke_Risk_in_Women_English_Infographic.png}
%     \caption{Examples of data narratives}
%     \label{fig:2DN}
% \end{figure*}

As indicated above, the corner stone of a \vero{DN} %data narrative (DN) 
is \vero{a} message, 
that associates characters with  measures.
Intuitively, a DN is an ordered set of messages and
will be manipulated based on the characters and measures
they deal with.

We list below simple queries that should be expressed over DNs.
Each one corresponds to an operation of the algebraic language
introduced in Section \ref{sec:language}. 

\begin{itemize}

\item \vero{Find DNs concerning some characters or measures, e.g., DNs about stroke deaths.} %What can be said about this character? About this measure? (
This \textit{selection} operation allows to find \vero{DNs} %narratives 
that satisfy a given condition. %) 

\item Retain from \vero{DNs} %narratives 
only the messages about \vero{some characters or measures, e.g., messages concerning Hispanic and Native American women.} %this character or measure (
This \textit{projection}\vero{-}like operation \vero{produces new DNs keeping} %keep from narratives 
only a subset of messages. %)

\item \vero{Concatenate messages of several DNs.} %Complete this DN with that one.
For example, produce a DN \vero{with all} %by concatenating the
messages of the DNs of Figure \ref{fig:2DN}.
\vero{This \textit{concatenation} operation allows gathering messages.} % of all DNs.} %%% TO WIN 1 LINE

% \item \vero{Combine two sets of DNs concatenating pairs of DNs.} 
% %\vero{For example, produce a DN by concatenating the messages of the DN of Figure \ref{fig:2DN}.}
% This %\textit{cross product} 
% operation over \vero{two} sets of DNs \vero{computes their Cartesian product.} %allows to concatenate DNs.  

\item \vero{Remove duplicate messages in DNs. This \textit{duplicate elimina-tion} %-like 
 operation only keeps one occurrence of each message in a DN.}

\item Synthesize groups of messages in \vero{DNs.} %a narrative. 
\vero{For instance, produce DNs aggregating messages about stroke cases and stroke deaths.}
%\vero{For instance, produce DNs with a synthesis of messages about stroke cases, followed by a synthesis of messages about stroke deaths.}
This \textit{group-aggregate}\vero{-}like operation groups messages in \vero{each DN} %the narrative 
using grouping conditions %, e.g., messages with related characters, and for each group, merges these messages 
\vero{and merges them} using built-in merging functions. %%% TO WIN SOME LINES 

\item Manipulate sets of \vero{DNs} %data narratives 
using the classical set operations (\textit{cross-product}, \textit{intersection}, 
\textit{union} and \textit{difference}). 

\item Change the plot of %the data narrative
\vero{DNs, for example, arranging messages according to some measures}. %(
This \textit{order by}\vero{-}like operation allows to modify the order of messages in the DNs. %)

\end{itemize}

Composing these operations enables to devise complex query expressions 
as will be seen in more details in Section \ref{sec:examples}.
We simply mention \vero{here} a few complex operations that we postulate
will be very useful in practice:

\begin{itemize}

    \item Connect the dots %\footnote{The expression is borrowed from \cite{DBLP:conf/kdd/ShahafG10}.} 
    between DNs, i.e., merge DNs that have characters, or characters and measures in common. This join\vero{-}like operation can be expressed with a combination of cross product, group-aggregate and selection.
    %What messages are common to these 2 narratives?
% compose messages from different narratives (to be merged with join, in the same way as the merge/fusion cube of Agrawal97) 
% \patrick{probably a shortcut for cross product followed by group-aggregate}
    
    \item Summarize a narrative 
    for a particular character (e.g., from women stroke to stroke, France to Europe, etc.). This roll-up\vero{-}like operation supposes to find in the DNDB narratives with messages having characters that generalize the particular character.
    It can be expressed using selection, cross product, projection, and group-aggregate.

% Compute DN by finding in the DNDB 
% messages summarizing messages of instance $I$,
% in the sense of the specialization relation 
% between messages.

\item Detail DNs for a particular character. This drill-down\vero{-}like operation can be seen as the inverse of the previous one and can be expressed with the same combination of selection, cross product, projection, and group-aggregate.
% Compute DN by finding in the DNDB 
% messages detailing messages of instance $I$,
% in the sense of the specialization relation 
% between messages.
% (drill-down) 

%Can be expressed using selection, cross product, projection, group.
\end{itemize}

\section{Logical data model}\label{sec:model}

This section presents the data model
of the logical framework. Again,
in this vision paper,
the  goal is not to define a thorough 
logical layer for DNs,
but instead to give the flavor 
of what this logical layer \vero{should be}
for an end-to-end DNMS.

\subsection{Data narrative components}

\paragraph{\textbf{Atomic concepts and relations}}

The atomic components of the model are
characters and  measures.
For instance, \vero{the} DN of Figure \ref{fig:2DN} (bottom)
includes character `black women' and measure `stroke death'. %"STROKE DEATHS BY ETHNICITY".
To keep things simple in this preliminary version
of the model, measure values and units
(e.g., 57/100000) are not part of the present
preliminary model. The semantics of the message
is given by a predicate connecting characters
and measures, in the spirit of semantic triples.

We also assume  binary relations between 
characters as 
the bases for the relations
between messages. Relations between characters
are at the core of relations
between findings
(see Figure \ref{fig:subCM}).
In data narration, it is common to 
use these relations as transitions between
episodes of the narrative's plot,
and it has been seen that most transitions
are of the following nature: specialisation, temporal or spatial
(cf. e.g.,  \cite{DBLP:journals/cgf/HullmanKL17}). 
More precisely, 
\begin{itemize}
    \item a  specialization relation, noted $c \prec c'$, 
    allows to classify characters in a hierarchy,
    for instance to indicate that black women is more specific than women,
    \item a  spatial relation, noted  $c \vdash c'$ to indicate
   that $c$ is in a spatial relation with $c'$. For instance \vero{Greece $\vdash$ France.}
   \item  a temporal relation, noted  $c \dashv c'$ to indicate
   that $c$ is in temporal relation with $c'$. For instance\vero{, in Europe, Spring $\dashv$ 2nd quarter}.
   \item we also assume a general similarity relation noted $c \approx c'$ to indicate that
    character $c$ is similar to character $c'$,
    for instance birth control pills is similar to abortion pills. 
\end{itemize}

% Without loss of generality, in what follows,
% we simply use $c R c'$ for one of such relation
% between characters $c$ and $c'$.

%\panos{PV: Maybe have the type of relationship as a modality? E.g., other types can be part-whole, alternative, sibling, ... Also the current ones are not clear, e.g., what does Greece $\vdash$ France mean? siblings? neighbors? with similar properties, e.g., like population?}

\paragraph{\textbf{Complex concepts}}

A message is a tuple associating characters
with measures. Formally, 
a message $m$ is a tuple \vero{$\langle C,V,P\rangle$} %$\langle C,M\rangle$
where $C$ is a set of characters, \vero{$V$} %$M$
is a set of measures and $P$ is a predicate\footnote{For the sake of consistency, $P$
is a singleton.}.
The simplest message is the empty message,
$\langle \emptyset,\emptyset, \emptyset\rangle$.

\begin{comment}
For instance, \vero{the} DN of Figure \ref{fig:2DN} \vero{bottom} %left
contains message \vero{$\langle C_1,V_1\rangle$} %$\langle C_1,M_1\rangle$
where $C_1=\{$Black, Hispanic, Native American, White, Asian$\}$
and \vero{$V_1=\{$STROKE DEATHS BY ETHNICITY$\}$.} %$M_1=\{$STROKE DEATHS BY ETHNICITY$\}$.
\end{comment}

\begin{figure}[ht]
    \centering
    \includegraphics[width=8cm]{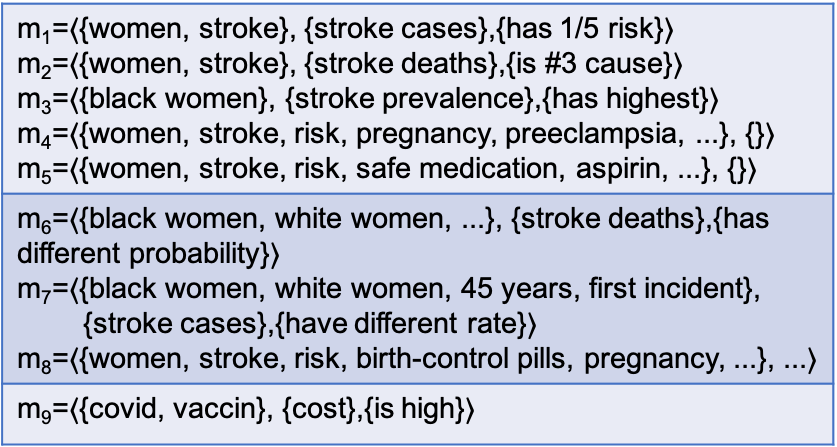}
    \caption{Messages of the running example}
    \label{fig:messages}
\end{figure}

\paragraph{Running example}
Consider the messages of Figure \ref{fig:messages}. Messages $m_1$ to $m_5$ and $m_6$ to $m_8$ are inspired from those of the DNs of Figure \ref{fig:2DN}, restricting to a subset of messages and simplifying many characters. Message $m_9$ is inspired from a DN about covid.  
\eop

\vspace{0.2cm}
Since findings can be related to one another
%(e.g., STROKE DEATHS BY ETHNICITY is more specific than DEATHS BY ETHNICITY), 
\vero{(e.g. findings about black women are more specific than those concerning all women),}
we consider that relations over characters also
applies to messages.
\vero{For example, among messages of Figure \ref{fig:messages}, $m_3$ is more general than $m_1$ regarding characters `black women' and `women'.}
%\vero{For example, there are three messages in the gray strip at the top of Figure \ref{fig:2DN}; the first one (1 in 5 women will have a stroke) can be seen as more general than the third one (Black women have the highest prevalence of stroke) regarding characters women and Black women.}
Given two messages $m, m'$ we consider
the transition relation between them
which can be one of: spatial, temporal, generalization, similarity.

Formally, for \vero{$m=\langle C,V,P\rangle$} %$m=\langle C,M\rangle$
and \vero{$m'=\langle C',V',P'\rangle$} %$m'=\langle C',M'\rangle$,
it is $m R m'$
if $\exists c \in C, c' \in C', c R c'$
and $R$ is one of 
$\prec, \approx, \vdash, \dashv$.

\subsection{Data model}

To be consistent with the conceptual description of DNs
of Figure \ref{fig:subCM}, we define a
 DN as a sequence of episodes. Each episode narrating
 a message, i.e., a tuple, a DN is formally defined
 as a tuple of tuples.
We distinguish its schema, which consists
in the number of messages (remember that
each message has the same structure),
from its instance.

Unless otherwise specified, all sets are infinite and countable.
Let $\mathcal{C}$ be a set of characters, %. Let 
$\mathcal{V}$ %be 
a set of measures,
$\mathcal{P}$ %be 
a set of predicates
\vero{and} %. Let 
$\mathcal{M}$ %be 
the set of messages 
%formed  from 
$2^\mathcal{C} \times  2^\mathcal{V} \times \mathcal{P}$.   %%% SMALL CHANGES TO WIN 2 LINES
Let 
%$\mathcal{MN}$ be a set of message names  and 
$\mathcal{H}$ be a set of DN names.
%\patrick{message names are introduced for
%compatibility with nested relations, see DN schema
%below. Check if it helps, or drop it}

\paragraph{Data narrative}

The schema of a DN of length $k$ (i.e., with $k$
messages) is a couple $\langle h,k\rangle$
where $h \in \mathcal{H}$ is the DN name.
%its structure, i.e., its name
%and its sort $\langle M_1,  \ldots, M_n \rangle $,
%where $M_1,  \ldots, M_n $ are message names.
%
%\paragraph{Data narrative instance}
%
A DN instance is a tuple of messages.

For the sake of readability, 
in what follows we will
consider a DN of length $k$
as an injective function from 
 $\mathcal{M}$
to $2^\mathbb{N}$.
For instance, the DN
$n=\langle m, m, m' \rangle $
can be seen as the function $n$
where $n(m)=\{1,2\}$ and $n(m')=\{3\}$.
We abuse notations and note
$m \in n$ if a message $m$
appears in  the DN $n$,
and $messages(n)$ the set of
messages of DN $n$.
%We also note $bags(n)$ the multiset of messages of a DN $n$.
% Note that the function is neither injective
% nor surjective since a message can appear
% at different positions of the DN,
% and there is only one message per position. 

\paragraph{Data Narrative Database (DNDB)}

A DNDB schema is a set of DN schemas and 
a DNDB instance is a set of DN instances.

\paragraph{Example}   %%% REORGANIZATION TO WIN 3 LINES
\vero{
Continuing the running example,
$I = \{n_1, n_2, n_3\}$,
is a DNDB instance that
organizes the
%the following DN instances organize 
messages of Figure \ref{fig:messages}: 
}

$n_1 = \langle m_1, m_2, m_3, m_4, m5 \rangle$;
$n_2 = \langle m_6, m_7, m_8 \rangle$;
$n_3 = \langle m_9 \rangle$.
\eop

\section{Data Narrative Manipulation Language (DNML)}\label{sec:language}

\vero{This section presents} %We present
an algebra for manipulating \vero{DNs.} %narratives. 
\vero{The focus is the description of the operators in their general form; more user-friendly constructors, specially for expressing conditions, will be discussed in Section \ref{sec:discussion}.}
All operators have the same signature in the sense
that they are applied over a DNDB instance
and output a DNDB instance.
Finally, note that DNs can be manipulated using
their schemas, i.e., their names and lengths. However,
it is preferable to also manipulate DNs with 
conditions over their messages
and this is why most operators rely on logical 
formulas for expressing these conditions.
In what follows, let $D$ be a DNDB and $I$, 
$I_1$ and $I_2$ be instances of $D$.

\subsection{Constants}

The first operation is 
the constant DN. Given a message 
$\langle C,V,P \rangle$,
it is simply: $\{ \langle C,V,P  \rangle \}$

% \centerline{
% $\{ \langle C,V  \rangle \}$.
% }

\subsection{Unary operators}

\paragraph{\textbf{Selection}} 

Selects the DNs in instance $I$ \vero{that satisfy a given} %based on some 
condition.

\centerline{
%$\sigma_{\varphi}(I)=\{n \in I | \varphi(n)$ is true $\}$
$\sigma_{\varphi}(I)=\{n \in I | \varphi \}$
}

where $\varphi$ is a logical formula to express selection conditions, for instance:
\begin{itemize}
    %\item $n.NAME=x$ (is of name $x$)
    \item $\exists m \in n, m=\langle C,V,P\rangle, c \in C$ (has character $c$)
    \item \vero{$\exists m \in n, m=\langle C,V,P\rangle, v \in V$ (has measure $v$)} %same with measure
     \item \vero{$\exists m \in n, m=\langle C,V,P\rangle, p \in P$ (has predicate $p$)}
    \item $\exists m \in n, m=\langle C,V,P\rangle, \exists c' \in C, c'Rc$ (has character in relation $R$ with $c$, %where   %%% TO WIN 1 LINE
    $R$ \vero{being} %is 
    one of the relations over characters)
    \item $\exists m, m' \in n, mRm'$ (has messages that are in relation $R$ where $R$ is one of the relations over messages)
%    \item $\forall m \in n, m\not=\langle \emptyset,\emptyset\rangle$ (has only non empty messages)
    \item $\forall m \in n, m=\langle \emptyset,\emptyset,\emptyset\rangle$ (has only  empty messages)
\end{itemize}

\paragraph{Example} The operation 
$\sigma_{\exists m \in n, m=\langle C,V,P\rangle, `stroke Deaths' \in V}(I)$ \\
looks for DNs about stroke deaths in instance $I$ of the running example. Its output is $\{n_1, n_2\}$. \eop

\begin{comment}
    
\faten{For example, assuming  a DNDB consisting of the 3 data narratives, which instance is $I=\{D_1,D_2,D_3\}$. The following operation outputs DNs of instance $I$ concerning characters similar to Stroke.
\centerline{
$\sigma_{\exists m \in n, m=\langle C,V\rangle, \exists c \in C, c \approx `Stroke'}(I)$
}
The result is: $I=\{D_1,D_2\}$ }

\vero{For example, the following operation outputs DNs of instance $I$ concerning characters similar to Migraine:}

\centerline{
\vero{
$\sigma_{\exists m \in n, m=\langle C,V\rangle, \exists c \in C, c \approx `Migraine'}(I)$
}
}

\end{comment}

\paragraph{\textbf{Projection}}

\vero{For each DN in I, keeps only the messages satisfying a condition.}
%Projects out undesired messages from DNs.

\centerline{
$\pi_{\varphi}(I)=\{ n|_{\{m \in n | \varphi\}}  | n \in I \}$
}

% where  $\varphi$ is a logical formula, for instance:
% \begin{itemize}
%     \item $ m=\langle C,V\rangle, c \in C$ (has character $c$)
%     \item $ m=\langle C,V\rangle, v \notin V$ (does not have measure $v$)
% %     \item same with measure
%     \item $ m=\langle C,V\rangle, \exists c' \in C, c'Rc$ (has character in relation $R$ with $c$, where $R$ is one of the relations over characters)
%     \item $ m=\langle C,V\rangle, \nexists c' \in C, c'Rc$ (does not have character in relation $R$ with $c$, where $R$ is one of the relations over characters)
%     \item $m\not= \langle \emptyset,\emptyset\rangle$ (is not the empty message)
% \end{itemize}

where  $\varphi$ is a logical formula,
similar to those used for the selection
\vero{and} $n|_{\{m \in n | \varphi\}}$ is
the restriction of DN $n$ to the set of 
messages satisfying $\varphi$.

% \patrick{could also be defined in the usual way
% with the message names, but the DNs have different
% sorts and arities.}

\paragraph{Example} The operation 
$\pi_{\exists m \in n, m=\langle C,V,P\rangle, `BlackWomen' \in C}(I)$
produce DNs $n_4, n_5$ and $n_6$, projecting a subset of messages of DNs in instance $I$, i.e., those containing Black women as characters.

$n_4 = \langle m_3 \rangle$;
$n_5 = \langle m_6, m_7 \rangle$;
$n_6 = \langle \rangle$.
 \eop

\begin{comment}
    
\faten{
For example, assuming a DN instance is $I=\{D_1,D_2\}$, the following operation outputs the DNs of instance I, restringing their messages to the ones measuring stroke deaths: \break
$\pi_{\exists m \in n, m=\langle C,V\rangle, `StrokeDeaths' \in C}(I)$ \break
The results is: D1 and D2 contains respectively  M12 and M21.}

\vero{For example, the following operation outputs DNs of instance I restring their messages to the ones measuring stroke deaths:}

\centerline{
\vero{
$\pi_{\exists m \in n, m=\langle C,V\rangle, `StrokeDeaths' \in C}(I)$
}
}

\end{comment}

\paragraph{\textbf{Duplicate-elimination}}

For \vero{each DN} %}all DNs 
in $I$, keeps only one occurrence of
each message.

\centerline{
 $\delta(I)=\{  n'  | n \in I, \forall m \in n, n'(m)=min(n(m)) \}$
}

\paragraph{\textbf{Group-aggregate}}

For \vero{each DN} %all DNs 
in $I$, groups the messages
using a grouping condition
and aggregates each group using a
specific aggregation function.

$\gamma_{\varphi_1, \ldots, \varphi_j, agg_1, \ldots, agg_j}(I)=$

\centerline{
$\{ \langle agg_1(\{m \in n | \varphi_1\}),
\ldots,
agg_j(\{m \in n | \varphi_j\})
\rangle | n \in I \}$
}

where the $\varphi_i$ are grouping conditions on the set of messages
of the DN. It is not requested that the grouping conditions
partition the set of messages, which allows one message
to appear in several groups.
The $agg_i$ are aggregation functions for merging 
a set of messages into one message.

\paragraph{Example} The operation
$\gamma_{BW,WW,A,A}(I)$, 
where $BW$ and $WW$ are conditions about characters `Black women' and `White women' and $A$ is an aggregation function computing the union of characters and measures in the input messages. The output DNs contain two messages, the former concerning Black women and the latter White women: 
%When applied to instance I of the running example, its output is:
%
$n_7 = \langle m_3, \langle \emptyset,\emptyset,\emptyset\rangle \rangle$; 
$n_8 = \langle A(m_6,m_7), A(m_6,m_7) \rangle$; \\
$n_9 = \langle \langle \emptyset,\emptyset,\emptyset\rangle, \langle \emptyset,\emptyset,\emptyset\rangle \rangle$.
 \eop

\paragraph{\textbf{Full group-aggregate}}

Another group-aggregate operation
allows to merge messages across DNs into one DN.

$\gamma^{across}_{\varphi_1, \ldots, \varphi_j, agg_1, \ldots, agg_j}(I)=$

\centerline{
$\{ \langle agg_1(\{m  | \varphi_1\}), \ldots, agg_j(\{m  | \varphi_j\}) \rangle
%| \exists n \in I, m \in n \}$
| m \in \bigcup_{n \in I} messsages(n)\}$
}

\paragraph{Example} The operation 
$\gamma^{across}_{BW,WW,A,A}(I)$, 
with $BW$, $WW$ and $A$ as in previous example, outputs

$n_{10} = \langle A(m_3,m_6,m_7), A(m_6,m_7) \rangle$
\eop 

\begin{comment}
For example, 
$\gamma^{across}_{\exists m \in n, m=\langle C,V\rangle, `stoke deaths' \in V,merge}(I)$ 
outputs $\{ \langle \langle \{women, stroke, black~women, white~women, \ldots \},$ $\{ stroke$ $deaths\} \rangle  \rangle  \}$
\eop 
\end{comment}

\paragraph{\textbf{Order by}}

Allows to change the order of messages in the DNs.

$\tau_{\varphi_1, \ldots, \varphi_J,sort_1,\ldots, sort_J}(I)=$

\centerline{
$\{ \langle sort_1(\{m \in n | \varphi_1\}),
\ldots,
sort_J(\{m \in n | \varphi_J\})
\rangle | n \in I \}$
}

where the  $\varphi_i$ are selection conditions
over messages (like the ones used e.g., for the selection operation)
and the $sort_i$ are sorting functions,
mapping a set of messages to a tuple of messages.
%with signature $2^{\mathcal{M}} \rightarrow \langle \mathcal{M}\rangle$.
% With this definition, it can easily be seen 
% that this order by operation is a particular case
% of the group-aggregate operation.
%\vero{TO REWIRTE TO FLATTEN MESSAGES TO ONE TUPLE. PERSONNALLY, I PREFERED THE SIMPLE VERSION.}

%\patrick{then some unnesting is required for closure }

\paragraph{\textbf{Concatenation}} 

Flattens all DNs in I into one DN, by concatenating all their messages.

\centerline{
$\chi(I)=\{ \langle n_1,...n_k \rangle  | n_i \in I\}$
}

% \paragraph{Example} The operation 
% $\chi(I)$ outputs DN 
% $n_{11} = \langle m_1,m_2,m_3,m_4,m_5,m_6,m_7,m_8,m_9 \rangle$
% \eop 

\subsection{Binary operators}

%\paragraph{\textbf{Cross product}} 

%The regular cross product  allows to concatenate DNs.

% \centerline{
% $I_1 \times I_2=\{ \langle n_1,n_2 \rangle  | n_1 \in I_1, n_2 \in I_2 \}$
% }

%\paragraph{\textbf{Union, intersection, difference}} 

The classical relational binary operations
$\times, \cup, \cap, 
\setminus$ 
have their standard set theoretic meaning.

% \faten{
% For example, assuming 2 DN instances, $I_1=\{D_2\}$ and $I_2=\{D_3\}$, the output of the cross product of I_1 and I_2 is: \break
%  $ D=\langle \break
%  M_2_1=\langle  \{blackwomen, whitewomen,asianwomen,etc.,covid,vaccin\},\{StrokeDeaths,Cost\} \rangle,\break
% M_2_2= \langle  \{blackwomen,whitewomen,45 years,first incident,covid,vaccin\},\{StrokeCases,Cost\}\rangle,\break 
%  M_2_3= \langle  \{women,stroke,risk,pregnant,etc.,covid,vaccin\},\{cost \}\rangle \rangle$   \break 
% }

\subsection{Properties}

We give here a few insights on 
the properties of the algebra, 
a complete study thereof
is part of our future work.

\paragraph{Closure}:
By definition, each operator defines a set of DNs from
a set of DNs.

\paragraph{Completeness}

From a given set of messages, 
all narratives can be obtained 
using the constant and cross-product
operators, which together form
the core of the algebra.

\paragraph{Minimality}

The minimal set of operators includes
constant, cross product, both group-aggregates,
order by,
union and difference.
All other operations can be expressed
from this set.
% In particular, selection, projection, duplicate elimination and order by can be 
% expressed using group-aggregate, \vero{SURE OF THIS?}
%and intersection can be expressed using difference.
%intersection can be expressed with Cross-product,
%selection and projection.
%Group-aggregate.
%
We can anticipate that some
expressions will be popular and deserve 
to be promoted as operators, like the join
in the relational algebra that is a shortcut
from cross product followed by selection.

\paragraph{Some properties of the operators}

Like in the case of the relational algebra,
the cross product is non commutative, admits one
absorbing element (the empty set)
and one neutral element (the empty DN, i.e., $\langle \rangle$).
% $\langle \langle \emptyset, \emptyset \rangle, 
% \ldots, \langle \emptyset, \emptyset \rangle
% \rangle$).
%
Set operations keep their usual properties.
Note that, because DNDB instances are sets of DNs
of different length, intersection cannot be
expressed by combining cross product, selection and projection, while it is the case for the relational algebra.

\section{Example}\label{sec:examples}

In this section, we give some examples
of useful and easy to express
queries in  DNML.
\vero{They are based} %All examples are given 
on the running example. %%% TO WIN 1 LINE

\paragraph{Comparing messages}

We first illustrate how to "join" DNs having  contradictory messages for characters
women and stroke, 
\begin{enumerate}
    \item select DNs about women and stroke:\\
     $I_1=\sigma_{\exists m \in n, m=\langle C,V,P\rangle, \{stroke,women\} \subseteq C }(I)$
    \item group by messages with women and stroke,  aggregate by keeping 
    messages that are contradictory:\\
    $I_2=\gamma^{across}_{\varphi,\neg \varphi ,check,drop}(I_1)$,     where \\
    $\varphi=\exists m \in n, m=\langle C,V,P\rangle, \{stroke,women\} \subseteq C $,
    $check$ is a function merging messages that are contradictory otherwise producing
    an empty message, and $drop$ is a function producing an empty message.
    \item project out empty messages:
    $\pi_{m\not= \langle \emptyset,\emptyset,\emptyset\rangle}(I_2)$
\end{enumerate}

\paragraph{Roll-up and drill-down}

%Let $I$ be a DNDB instance and $n$ be a DN. 
Assume we want to see DNs about 
'black women, stroke' and then to
"roll-up"  from
'black women'.

\begin{enumerate}
    \item select the DNs with characters
    'black women','stroke'\\
    $I_1=\sigma_{\exists m \in n, m=\langle C,V,P\rangle, \{'black~women','stroke'\} \subseteq C }(I)$
    \item select the DNs with characters more general than 'black women'\\
    $I_2=\sigma_{\exists m \in n, m=\langle C,V,P\rangle, 'black~women'\prec x, x \in  C }(I)$
    \item compute the cross product of found DNs:\\
    $I_3=I_1\times I_2$
    \item group by messages with black women and one more general character,  aggregate by merging 
    messages:\\
    $I_4=\gamma_{\varphi,\neg \varphi ,merge,drop}(I_3)$,     where \\
    $\varphi=\exists m \in n, m=\langle C,V,P\rangle, \{'black~women'\} \in C \vee 'black~women'\prec x, x \in C $,
    $merge$ is a function merging messages, and $drop$ is a function producing an empty message.
    \item project out empty messages:
    $\pi_{m\not= \langle \emptyset,\emptyset,\emptyset\rangle}(I_4)$
\end{enumerate}

\section{Related work}\label{sec:rw}

\paragraph{Data narrative modeling}

Calegari et al. \cite{DBLP:conf/clei/Calegari22}
proposed a narrative metamodel based on the conceptual model of \cite{DBLP:conf/er/OutaFMPV20}, to provide abstract models to data narratives. They explored the definition of model transformation for converting narrative models into  HTML or a Jupyter computational notebook.
%This work proposed no manipulation of the data narrative; it was merely a model-to-text transformation to HTML and Jupiter.\\
%
Zhang et al.\cite{framework_datastory_2022} proposed a framework for creating data storytelling applications from three major perspectives: concept, component, and procedure with the absence of any logical means.
Bach et al. \cite{bach2018narrative}  introduces narrative design patterns, defined as "a low-level narrative device that serves a specific intent". A pattern can be used individually or in combination with others to give form to a story. Five major patterns of group are identified: argumentation, flow, framing, emotion, engagement. For example, if the intent of the data narrator is to \textit{persuade} and \textit{convince} audience, he can use one of the following patterns: compare, concretize, and repetition. Importantly, these patterns are not specifically related to a visualization or interaction medium.

Many approaches exist for describing 
DN crafting, mostly describing
an essentially manual process
\cite{chen/storysynthesis,DBLP:journals/cga/LeeRIC15,DBLP:conf/ivapp/DuangphummetR21}, 
while others proposing approaches for
automatically generating simple data narratives
 \cite{DBLP:journals/tvcg/WangSZCXMZ20,DBLP:journals/tvcg/ShiXSSC21,DBLP:journals/cgf/ShiSXLGC21}.
In all cases, no manipulation language was specifically 
proposed for manipulating DNs.
%based on information unit structuring, associated computational technologies, methods emerging from serious 

\paragraph{Languages for DN} 

A DNDB is a set of tuples of tuples of sets, i.e.,
a  form of nested relation,
albeit with tuples of different lengths.
This means that 
the proposed language, DNML, is likely to be expressed
in the nested relational algebra (NRA)
\cite{DBLP:books/aw/AbiteboulHV95}.
However, due to the relatively  simple structure of DNs, some of NRA's operations are not needed
(e.g., nesting/unnesting, powerset).

% languages for sequence, 
% \cite{DBLP:journals/tkde/MeccaB01}
% languages for JSON (https://arxiv.org/abs/2006.04277)
% Note that in our case infinite answers are not possible.

Many languages or primitives were proposed
to express data exploration sessions \cite{DBLP:journals/is/VassiliadisMR19,DBLP:conf/sigmod/ElMS20,DBLP:journals/pvldb/YoungmannAP22}.
While relevant for understanding the logic 
under the discoveries of finding,
these languages are not adapted the 
manipulation of messages and are not
devised as an algebra.
Some are not even meant to be used by humans, 
since, e.g., in \cite{DBLP:conf/sigmod/ElMS20}, primitives
are used to generate exploration sessions
through reinformcement learning.

% A set of novel intentional OLAP operators of a high-level language, 
% which express the user’s need for results, is proposed in \cite{}. For example, the \textit{describe} operator is used when the user’s intention is to know something more about a set of facts; \textit{predict} when the user aims at predicting about future values; \textit{explain} when the user aims at understanding the cause of a phenomenon; \textit{assess} when the user aims at comparing and bringing in more data to the ongoing analysis; and \textit{suggest} when the user aims at asking for guidance.

%\patrick{Check also Manolescu's work on fake news.}

%Data narrative: Blount and narrative patterns:
%https://www.scitepress.org/Papers/2020/101216/101216.pdf
%(see esp. section 2.3)

\section{Discussion}\label{sec:discussion}

We close this paper by discussing
some of the main challenges raised by the 
development of end-to-end DNMS.

\paragraph{Modeling the complexity of data narration}

The logical layer proposed in this paper
only covers a small portion of 
data narration, i.e., the complex process that
goes from data exploration to the visual
presentation of messages.
In particular, to account for the complexity
of the process, all the concepts
and relations present in the model of \cite{DBLP:conf/er/OutaFMPV20}
should have a counterpart in the logical layer.

For instance, the provenance of messages
(how findings in the dataset were discovered),
and the meaning carried to the reader
should be logically modeled.
For provenance, messages should be linked to a collection of findings,
independently of the form of findings, as well as to the queries
the findings are results of.
%This is as generic as it gets, independently of the form of findings
%(i.e., even if the data are in a database, the findings are results of queries, and modeled as stand-alone objects, so they mask the underlying data)
For the semantics, one can assume extending the message predicates 
with user-intuitive semantics,
which requires
including measure values, units,
quantification, etc.
%the participation of characters and measures as their arguments,
%possibly with other parameters too.
%
A challenge will be to cover all the data narration process 
while keeping the model and language
simple enough.

\paragraph{Data model}

As noted above, the data model proposed above
is very close to RA. In fact, if the order of messages is not 
included in the data model, DNs could  simply be defined
as sets of messages, i.e., relations of arity 2.
However, this simplicity would oblige to code the
complexity of DN, leading to unnatural queries.
Besides, as explained in the previous 
paragraph, this model is meant to be extended to
cover the complete conceptual model of \cite{DBLP:conf/er/OutaFMPV20}.

More semantics should be added in the different layers
of the conceptual model of \cite{DBLP:conf/er/OutaFMPV20}.
For instance, at the message level, this can be achieved
by using  predicates having semantics known to the reader,
like in \cite{DBLP:journals/tvcg/WangSZCXMZ20},
or by modeling the relations between characters using
e.g., property graphs.
At the data exploration layer,  modeling findings can be done
by relating characters and measures
in the spirit of \cite{DBLP:journals/is/AnadiotisBCGHMM22}.
Semantics can also be added by modeling the intentions
of the narrator using abstract primitives like the ones
of \cite{DBLP:journals/is/VassiliadisMR19} 
or narrative patterns of Bach et al. \cite{bach2018narrative}.

\paragraph{Manipulation language}

As mentioned above, extending DNML will be needed to 
account for the complexity of the data narration process. 
An eye should be kept on query languages for sequences
\cite{DBLP:journals/tkde/MeccaB01} and query
languages for the Semantic Web \cite{DBLP:journals/tods/ArenasGP18}. 
A challenge will be to keep the formalism simple enough
for ensuring its adoption by the narrators, analysts or
data enthusiasts. 

This can be achieved by, e.g.,  (i) devising operations
summarizing complex expressions that
are useful in practice,
in the spirit of join for RA,
(ii) devising a SQL-like language for DNs, 
(iii) using built in predicates
for the logical formulas used in the operations 
(e.g., selection, projection, group-aggregate).
In this last case, for instance, 
predicate  $hasChar(c)$ could be used
to express that DN $n$ has character $c$, i.e., logical formula
 $\exists m \in n, m=\langle C,V,P\rangle, c \in C$.

\paragraph{RDBMS like stack}

On the long term, the challenge will be to 
implement a DNMS on the model of the
successful achievements of RDBMSs,
notably, a clear distinction of
conceptual, logical, physical layers
with intuitive mappings between
the objects of different layers,
loading facilities for populating 
the DNDB from existing DNs,
data organization at the physical layer, including 
specific index mechanisms (e.g., inspired by 
information retrieval techniques), 
optimizations at the logical and physical layer,
etc.

\bibliographystyle{ACM-Reference-Format}
\bibliography{bibStory,biblio,recentBib,suppBib,basic}

\end{document}